# Pressure-induced non-monotonic crossover of steady relaxation dynamics in a metallic glass

Xin Zhang,[1,5] Hongbo Lou,[1,5] Beatrice Ruta,[2,*] Yuriy Chushkin,[3] Federico Zontone,[3] Shubin Li,[2] Dazhe Xu,[1] Tao Liang,[1] Zhidan Zeng,[1] Ho-kwang Mao,[1,4,*] Qiaoshi Zeng[1,4,*]

[1.] Center for High Pressure Science and Technology Advanced Research, Pudong, Shanghai 201203, P.R. China

[2.] Université Lyon, Université Claude Bernard Lyon 1, CNRS, Institut Lumière Matière, Campus LyonTech - La Doua, F-69622, Lyon, France

[3.] ESRF- The European Synchrotron, CS 40220, 38043 Grenoble, France

[4.] Shanghai Key Laboratory of Material Frontiers Research in Extreme Environments (MFree), Shanghai Advanced Research in Physical Sciences (SHARPS), Pudong, Shanghai 201203, P.R. China

[5] These authors contributed equally to this work.

* To whom correspondence should be addressed: zengqs@hpstar.ac.cn, beatrice.ruta@univ-lyon1.fr, or maohk@hpstar.ac.cn

## Abstract

Relaxation dynamics, as a key to understand glass formation and glassy properties, remains an elusive and challenging issue in condensed matter physics. In this work, *in situ* high-pressure synchrotron high-energy x-ray photon correlation spectroscopy has been developed to probe the atomic-scale relaxation dynamics of a cerium-based metallic glass during compression. Although the sample density continuously increases, the collective atomic motion initially slows down as generally expected and then counter-intuitively accelerates with further compression (density increase), showing an unusual non-monotonic pressure-induced steady relaxation dynamics crossover at ~3 GPa. Furthermore, by combining *in situ* high-pressure synchrotron x-ray diffraction, the relaxation dynamics anomaly is evidenced to closely correlate with the dramatic changes in local atomic structures during compression, rather than monotonically scaling with either sample density or overall stress level. These findings could provide new insight into relaxation dynamics and their relationship with local atomic structures of glasses.

**Classification:** Physical Science, Applied Physical Sciences

**Main text**

Below glass transition temperatures, $T_g$, glasses fall out of equilibrium (breakdown of ergodicity) due to their slow dynamics with respect to the experimental timescales; hence, dynamic relaxation into equilibrium states naturally and inevitably occurs all the time, considerably affecting the properties of glasses (1, 2). Therefore, enduring theoretical and experimental efforts have been devoted to understanding the glass relaxation dynamics over decades, especially their structural basis, i.e., the relationship between dynamic relaxation and the atomic structures of glasses (3).

Although lacking long-range translational symmetry, the local structure of seemingly disordered glasses still shows identifiable order, which may play a critical role in determining their structures and properties (4). Therefore, using the local structural order (e.g., short-range order, SRO, and medium-range order, MRO) as structural indicators to correlate with dynamic relaxation phenomena has attracted considerable research interests (5-13). In theories and simulations of glasses and glass formation, many local structural quantities, such as cooperatively rearrangement region (5), locally favored structures (6-8, 11), and 'softness' (9, 10), are proposed, which reveal complex scenarios of relaxation or aging at the atomic scales. However, it is rather difficult to verify these scenarios experimentally even in simple atomic glasses without complex intramolecular structure and freedom, like metallic glasses (MGs) (14).

Previous experimental investigations on the relaxation dynamics of glasses mainly rely on the macro-relaxation behavior of a given physical property, such as the dielectric constant or loss modulus (14). However, because these physical properties reflect the relaxation of the entire system, it is challenging to correlate them with local glass structures directly. Recent technical improvements in x-ray photon correlation spectroscopy (XPCS) have enabled measurements of collective particle motions of glassy samples with a high resolution in reciprocal space and broad coverage in the time scale (15). By utilizing this approach, various microscopic dynamic processes have been explored, including atomic diffusion and phase transformations (16-18), relaxation and aging dynamics in colloidal suspensions or gels (19, 20), ferrofluids, ferromagnets and ferroelectrics (21, 22), and MGs (23-28). In the case of MGs, as a simple atomic glass system, many interesting and unusual phenomena that are otherwise inaccessible have been observed by wide-angle high-energy XPCS, which sheds new light on the glass relaxation dynamics. For example, the compressed exponential relaxation revealed in atomic glasses (23, 24), anti-aging phenomenon observed in an ultrastable glass (25), and abrupt intermittent fluctuations (26) revealing the complexity of local relaxation events and the existence of a multi-step relaxation process in MGs (26-28). On the other hand, conventional tuning approaches, such as plastic deformation (29) and thermal cycling (30), can effectively alter the energy states and corresponding relaxation dynamics in MGs. However, the change in atomic structures is subtle or not noticeable (31), suggesting likely decoupling of the structure and relaxation behavior in those glasses.

To overcome the challenges encountered in previous experiments for addressing the relationship between relaxation dynamics and glass structures, we deliberately introduced dramatic structural tuning of MGs by applying high pressure while simultaneously probing the atomic-scale relaxation behavior by developing *in situ* high-pressure synchrotron high-energy XPCS (32). As a fundamental parameter dictating the states of matter, pressure is a powerful tool that can readily change the density of MGs by tens of percentage during compression (above the GPa level) and also could tune the atomic structure significantly, even with pressure-induced polyamorphic phase transitions in MGs (33). Actually, besides cooling (changing both thermal energy and volume), a liquid can be also vitrified by increasing pressure (only changing volume) under isothermal conditions. Understanding pressure dependence of glass properties is crucial to provide the complete physical description of glassy phenomenon. Under pressure, the previously proposed dynamics-sensitive structural indicators and critical parameters, such as the local atomic packing (cluster distribution), free volume, diffusion, internal stress level of an MG, etc, can be gradually and effectively altered but without changes of thermal energy, offering a unique and quantitative opportunity to clarify the correlation between relaxation dynamics and atomic structural features in glasses.

Specifically, in this work, by combining synchrotron wide-angle high-energy XPCS with a diamond anvil cell (DAC), we developed the *in situ* high-pressure and high-energy XPCS technique and performed it on a cerium-based MG. We show that the microscopic structural relaxation process is not monotonically altered as typically expected with the pressure-induced density and stress increase. Instead, it shows an anomalous dynamics crossover at ~3 GPa. The dynamics below ~3 GPa follows a normal pressure-induced slowing down while it accelerates rapidly above ~3 GPa. The coincidence of the dramatic local structural change revealed at ~3 GPa by *in situ* high-pressure synchrotron x-ray diffraction (XRD) measurement indicates a coupling with the dynamics anomaly and dramatic structural changes.

A polyamorphous cerium-based MG, $Ce_{68}Al_{10}Cu_{20}Co_2$, was chosen as the model system for the current study due to its well-known pressure-sensitive atomic structures and relatively low (near room temperature) glass transition temperature ($T_g$ ~360 K) (34, 35), which largely facilitates the XPCS measurement of glass dynamics in a feasible time scale (a few thousand seconds) at room temperature (~0.86 $T_g$ at 0 GPa) with obvious structural tuning at moderate pressures. Details about the MG sample synthesis and characterization can be found in Methods and Supporting information (SI) Fig. S1. The *in situ* high-pressure XPCS experiments were performed using a symmetric DAC and a partially coherent x-ray beam with a high incident energy of 21.0 keV at beamline ID10 of the European Synchrotron Radiation Facility (ESRF), Grenoble, France. During the high-pressure XPCS experiments, special attention was paid to the stability of the sample position and pressure (fluctuation < 0.01 GPa) since it is crucial for a reliable XPCS measurement (32). More details can be found in Methods and SI Fig. S2.

Figures 1a-e show the time evolution of the atomic-scale dynamics at different pressures captured by XPCS at $q_p$ (the wave vector of the principal diffraction peak position) with the two-time intensity correlation function (TTCF). This function maps the instantaneous correlation between the intensity scattered at subsequent delay times

$t_1$ and $t_2$ and can be considered as a time-resolved version of the more common time average intensity auto-correlation functions, $g_2(q, t)$. Since the pressure in the DAC is manually increased and the required waiting time for stabilizing pressure is pressure and stress rate dependent, this work only focuses on the final stationary regime of relaxation, as shown in Fig. 1, without including the initial aging process right after each pressure increase (aging is strain-rate and waiting-time dependent (26-28)). At each pressure, the width of the reddish diagonal contour, proportional to the relaxation time, remains almost constant in the measuring time window shown in Figs. 1a-e, indicating that steady dynamics were measured (15). It should be noted that slight aging still presents at 2.9 GPa, a signature of the persistent structural changes at this pressure, even at around three hours after changing pressure. Although this aging may underestimate the real steady relaxation time, it does not conceal the unusual evolution of the dynamics with further compression to higher pressures. For compression below ~3 GPa, the atomic motion slows down with increasing pressure. Then, it dramatically accelerates at pressures higher than ~3 GPa, as signaled by the increasingly thinner diagonal contours of the TTCFs at higher pressures.

We average each TTCF plot over the entire measuring time window to get quantitative information on the relaxation process. The time average intensity auto-correlation functions, $g_2(q, t)$, are shown in Fig. 1f. These functions were then fitted to a modified Kohlrausch-Williams-Watts (KWW) function: $g_2(q_p, t) = a + c[\exp(-2(t/\tau)^\beta)]$(20), where $\tau$ is the relaxation time, $\beta$ is the shape parameter, $a$ is the baseline involving the influence from the high-pressure environment (e.g., the diamond anvils and the pressure medium, see more details in Methods and SI text I), and $c = \gamma f_{D-W}^2$ is the product between the experimental contrast $\gamma$ and the square of the nonergodicity parameter of the glass. The $g_2$ function describes the decay of density fluctuations and provides quantitative information on the relaxation dynamics at the atomic scale. The pressure-dependent fitting parameters are shown in Figs. 2a-c. Details about the data analysis are provided in Methods and SI text I. In agreement with the TTCFs, the fitted relaxation time, $\tau$, does not change monotonically with pressure, but peaks at ~3 GPa, then decreases with further compression. In the previously reported temperature-variant XPCS experiments on MGs, $\tau$ usually changes monotonously with temperature during steady relaxation (23), although more heterogeneous dynamics have been reported during aging or stress relaxation (26, 28). According to the pressure dependence of $\tau$ below 3 GPa, although there are only three data points, the apparent activation volume according to the framework of the transition state theory, $\Delta V_a \sim 10$ Å$^3$ can be roughly estimated according to the relation (36), $\Delta V_a = RT (\mathrm{d} \ln \tau / \mathrm{d} P)_T$, where $R$ is the gas constant, $T$ is temperature, and $P$ is pressure. This roughly obtained activation volume value for relaxation of MGs under high pressures is comparable to the average atomic volume in the MGs, which is also close to those determined by dynamic mechanical analysis under tension stress on La-, Pd-, and Zr-based MGs (37). In contrast, the negative pressure dependence of steady relaxation time above 3 GPa (Fig. 2a) suggests a "negative" activation volume of ~-10 Å$^3$, which, in contrast, is quite unusual and cannot be explained by any classical models. Unlike the anomalous behavior of relaxation time, the shape parameter, $\beta$, keeps almost constant at ~1.8 in the studied pressure range (Fig. 2b). The contrast, $c$, slightly increases with pressure (Fig. 2c). The behavior of these two parameters will be discussed later.

Due to the close critical pressures, it is likely that the apparent dynamics crossover at ~3 GPa revealed by $\tau$ during compression might be associated with the structural polyamorphic transition from low-density amorphous (LDA) to high-density amorphous (HAD) states at a few GPa in many Ce-bearing MG (38, 39). The atomic structure evolution of the Ce$_{68}$Al$_{10}$Cu$_{20}$Co$_2$ MG with increasing pressure was further determined by *in situ* high-pressure XRD following the same compression protocol of the XPCS experiments to address the underlying mechanism of the dynamics anomaly and to clarify the structural transition scenario. According to the *in situ* high-pressure XRD data, the structure factor, $S(q)$, at different pressures and the corresponding reduced pair distribution function (PDF) curves in real space can be derived and are shown in Fig. 3 and 4, respectively. The inverse of the principal peak position, $2\pi/q_p$, versus pressure, as shown in Fig. 3c, indicates that density increases monotonously over the entire pressure range but with accelerated rates through the polyamorphic transition zone (34), whose inflection point nearly coincides with the turning pressure of $\tau$.

Moreover, pronounced changes can be observed in the first peak of $G(r)$ ($R1$, the first atomic shell) during compression. For quantitative analysis, the first peak of $G(r)$, $R1$, can be deconvoluted into two subpeaks, denoted as $r_{11}$ and $r_{12}$ (see Fig. 4a). Based on a simple hard-sphere model, the $r_{11}$ mainly consists of contribution from the Ce-Cu atomic pairs, while the $r_{12}$ is dominated by the contribution from Ce-Ce atomic pairs at ambient conditions (see details in SI text II). During compression of a Ce-bearing MG, a double-well potential of the Ce-Ce pairs will emerge, more and more "big" Ce atoms with localized 4$f$ electrons will gradually transform into "small" Ce atoms with delocalization of 4$f$ electrons accompanied by dramatic atomic volume collapse (38, 39). Therefore, more and more Ce atoms initially in the $r_{12}$ sub-shell will be squeezed into the $r_{11}$ sub-shell (redistribution of atoms in SRO), leading to a remarkable shape change ($r_{11}$ peak intensity increases at the expense of the $r_{12}$ peak intensity) of the first peak of $G(r)$ during compression. Figs. 4b-d show the variation of sub-peak positions ($r_{11}$, $r_{12}$) and the peak area of the first peak ($A$) as a function of pressure, which reflect the changes in the average bond length and coordination numbers in the two sub-shells of the nearest neighbor environment. The monotonic increase of the peak area (Fig.4b) indicates the typical pressure-induced densification, consistent with simulation results in other Ce-based MGs during compression (38). Fitting details of the two sub-peaks can be found in SI Fig.S5. Similar crossover phenomena can be observed in both Figs. 4c and 4d at ~3 GPa, in good agreement with the crossover of the relaxation time revealed by XPCS. Dramatic variation in SRO will inevitably extend into MRO (connection between SRO clusters) and thus to the corresponding collective atomic motion. Details about the extended structural changes in MRO and $G(r)$ data in a broader $r$ range are provided in SI text III.

The coincidence of the critical pressures for the dynamic crossover and the structural polyamorphic transition indicates an explicit correlation between them. At first glance, a transition from the LDA to HDA phases would naturally lead to an alternation of the glassy properties, including relaxation dynamics. However, given that the monotonic density increases with pressure through the polyamorphic transition, relaxation dynamics is expected to simply slow down according to the classical "free volume" model (40). A simple structural basis for micro-relaxation in MGs is escaping from a "cage" (an atom jumps away from its surrounding atoms). If the local atomic structure is "simply" compressed or shrunk under pressure, the "cage effect" is expected to be enhanced. Therefore, we should observe an accelerated slowing down of dynamics through the polyamorphic transition zone with faster density increase. However, against this expectation, a sharp turn around to accelerated dynamics accompanies the polyamorphic transition. The non-monotonic evolution of the dynamics behavior with respect to monotonic density increase is unexpected and counter-intuitive. According to the prominent changes in SRO over the same pressure range, as shown in Fig. 4c and 4d, it is suggested that the relaxation behavior revealed by XPCS might relies more on specific SRO details and their influence at extended length scale (rather than simply scaling with overall density). In different MG systems, dominant SROs are different, so their roles in relaxation behavior may vary from one to another. For example, the characteristic atomic cluster in $Cu_{64}Zr_{36}$ MG is full icosahedra (Voronoi index <0, 0, 12, 0>) (41), but in $Mg_{65}Cu_{25}Y_{10}$ MG, bicapped square antiprisms (<0, 2, 8, 0>) and tricapped trigonal prisms (<0, 3, 6, 0>) are dominant clusters (42). Evidence of a strict relation between local atomic structures and collective motion has been recently reported for a supercooled metallic liquid (43).

The LDA and HDA states of the $Ce_{68}Al_{10}Cu_{20}Co_2$ MG differ significantly in their SRO, which supports the important role of local atomic structure in affecting glass relaxation dynamics. It has to be noted that the pressure-induced polyamorphism is caused by the change of electronic structure in the Ce atoms (39). The electronic structure transition is directly coupled with the atomic rearrangement during the polyamorphic transition. During the polyamorphic transition, a double-well feature emerges in the Ce-Ce interatomic potential with a relatively small barrier between the two potential wells, which replaces the regular single-well potential (38, 39). According to a recent simulation work dealing with pressure-induced structural rejuvenation (44), this emerging two-state feature in potential could lead to structural rejuvenation and therefore accelerated relaxation dynamics. Although the pressures studied in this work are below 10 GPa because of the hydrostatic pressure limit of the 4:1 methanol-ethanol pressure medium (45), the relaxation dynamics are expected to slow down again when the polyamorphic transition completes at high enough pressures. But relaxation dynamics also could become even more complex if extreme pressure-induced structural rearrangement dominates. However, all these speculations need further experimental efforts to clarify in the future.

In addition to the anomaly of relaxation time, the structure dependence of other fitting parameters is also interesting and inspiring. The shape parameter, $\beta$, reflects the nature of the relaxation mode. A compressed decay ($\beta > 1$) is common in MGs and usually suggests a collective ballistic motion of atoms due to internal stress (23). In previous studies, $\beta$ exhibits temperature independence until heated to the glass transitions region, which suggests its density (thermal expansion) independence but high sensitivity to the frustrations or ordering of local structures (46). Moreover, the origin of compressed decay ($\beta > 1$) is proposed to be structural, e.g., microscopic collapse (47, 48), long-range microscopic spatial correlations (49), clustering of icosahedra (24), or elastic frustration (50) in many hard or soft materials. In this work, $\beta$ remains constant within error bars, suggesting its independence from changes in the SRO and density of a glass and also agrees with recent results on other MGs at the glass transition (43). In contrast, another parameter, $c$, is found to couple with glass structures. This parameter is related to the nonergodicity factor of the glass, which is associated with the elasticity of the glass and could provide indirect information on the occurrence of faster secondary relaxation processes (51). In the macroscopic limit, the nonergodicity level is related to the elastic properties of the glass, and it usually increases with increasing stiffness in the material (52), which is consistent with our observation, as shown in Fig. 2c. Due to the relatively large fitting uncertainty of $c$, a crossover at ~3 GPa in the pressure dependence of $c$ is not conclusive, which needs future efforts to clarify as well.

In this work, the XPCS signals are only collected at the principal diffraction peak position, $q_p$, in reciprocal space, which reflect collective atomic motion, but whether it corresponds to a well-defined length scale in real space is still controversial (53). Further exploration with XPCS data collected over a wide wave vector range ($q$-dependence of dynamics) could help to address this issue. It should be noted that the pressure-induced dynamics anomaly should not be limited to the specific composition (the Ce-based MG just facilitates the experiments with relatively low pressures to observe the dynamics anomaly) since pressure-induced structural rearrangement could be general in multicomponent MGs at high enough pressures (44). Generally, the pressure-induced changes of $\tau$ (probably $c$ as well) could correspond to the pressure-induced evolution of the metabasin in the energy basin of MGs from the potential energy landscape point of view (54). Therefore, applying pressure to MGs might lead to an evolution of local relaxation behavior (tuning the roughness of the potential energy basins), offering a promising tool for the experimental exploration of the long-sought Gardner transition in atomic glasses as well (55).

In summary, by combining *in situ* high-energy high-pressure XPCS and XRD techniques, an anomalous relaxation behavior at the atomic scale was discovered in the $Ce_{68}Al_{10}Cu_{20}Co_2$ MG, which shows a steady relaxation dynamic crossover at ~3 GPa with opposite pressure dependence below and above the critical pressure. Pressure-induced monotonic crossover has been previously reported in ionic liquids with enhanced slowing down of relaxation dynamics, which suggests a liquid-to-liquid transition (56). In contrast, it is quite unusual to observe a non-monotonic crossover from normal pressure-induced slowing down turning around to acceleration in relaxation dynamics. By analyzing the detailed structural information in $S(q)$ and $G(r)$ data derived from XRD, this relaxation anomaly is suggested to couple with the dramatic modification in local structures induced by the pressure-induced structural polyamorphic transition. Our findings demonstrate that the relaxation dynamics do not simply scale with overall glass density (free volume content (57)) or overall stress level.

Instead, there is a strong correlation between the variation of relaxation dynamics and local atomic structure (potential energy landscape) details of MGs. The experimental methods successfully demonstrated in this work highlight the feasibility and effectivity of involving pressure as a powerful parameter to explore the structural relaxation dynamics at the atomic level and its relationship with atomic structures in MGs by employing *in situ* high-pressure high-energy XPCS. This technique could enable the atomic-level dynamics exploration with pressures up to hundreds of GPa without limitation in types of materials (different physical structures and chemical natures) (32).

**Materials and Methods**

**Sample preparation.** $Ce_{68}Al_{10}Cu_{20}Co_2$ (nominal composition) MG ribbons were prepared by arc-melting pure elements (>99.9 at.%) and melt-spinning master ingots under a high-purity argon atmosphere. Their glass nature was confirmed by x-ray diffraction (XRD) and differential scanning calorimetry (DSC) (SI Fig. S1). Initialization was performed to eliminate the difference in thermal history among the as-prepared ribbon samples. Specifically, all the samples were heated up to 360 K with a heating rate of 20 K/min and then cooled down to room temperature with a ballistic cooling rate.

***In situ* high-pressure XPCS.** *In situ* high-pressure XPCS experiments were performed using a symmetric DAC at beamline ID10 of ESRF, Grenoble, France. We used a partially coherent x-ray beam with a high incident energy of 21.0 keV to effectively penetrate the two diamond anvils (~1.2 mm each, ~2.4 mm in total thickness) with the transmission of ~75.4%. The x-ray beam size of ~10 μm × 10 μm defined by rollerblade slits was smaller than the sample size, which helps to avoid signals from the sample edges (with plastic deformation during sample cutting) and the crystalline gasket. The speckle patterns were collected in a wide-angle transmission geometry by a CdTe Maxipix detector (512 × 512 pixels, pixel size of ~ 55 μm) placed downstream of the DAC with $2\theta \approx 16°$ at 0 GPa. The accurate $2\theta$ value is pressure dependent and was determined by scanning the detector position along $2\theta$ to find the intensity maximum at each pressure as the first principal diffraction peak position ($q_P$) of the $Ce_{68}Al_{10}Cu_{20}Co_2$ MG. Sets of ~1500 images were taken with 5 s exposure time per frame and analyzed according to the following equation:

$$g_2(q,t) = \frac{\langle\langle I(q,t_1)I(q,t_2)\rangle_p\rangle}{\langle\langle I(q,t_1)\rangle_p\rangle\langle\langle I(q,t_2)\rangle_p\rangle} \tag{1}$$

Where $<\cdots>_p$ denotes an ensemble average over all the pixels of the detector, and $<\cdots>$ is the temporal average.

The high-pressure samples were cut from the thermally initialized MG ribbons with an appropriate size (~90 × 90 × 20 μm³) to fit the sample chamber in the DAC. Tiny ruby balls were used as pressure calibrants loaded along with the MG sample. The culet size of the diamond anvils is ~300 μm. The gasket is T301 stainless steel, which was pre-indented to ~20 GPa in the DAC. The sample chamber is a hole (~150 μm in diameter) drilled by laser in the indent center of the pre-indented gasket. The 4:1 methanol-ethanol (4:1 M-E) mixture was loaded into the DAC as the pressure-transmitting medium, which can provide a hydrostatic pressure condition with no first-order phase transition below ~10 GPa (45). A schematic illustration of the experimental setup and sample image loaded in the DAC are shown in SI Fig. S2.

As the development and demonstration of *in situ* high-pressure, high-energy XPCS technique, many experimental details were carefully evaluated to ensure the reliability of the XPCS data. Special attention was paid to the sample stability since it is critical for a reliable XPCS measurement. Firstly, the DAC was firmly fixed onto the sample stage, and its spatial stability was checked by sample position scanning before and after each XPCS measurement. The pressure was increased manually by minimal steps up to each target pressure. Before performing XPCS experiments, the pressure inside the DAC was continuously monitored by the ruby fluorescence peak position to ensure a constant pressure value with time-dependent pressure fluctuations less than 0.01 GPa. Approximately three hours of waiting time (between each pressure increase and final XPCS data collection) were taken to stabilize the sample chamber pressure to ensure a steady relaxation signal detection. After each XPCS experiment, the pressure inside the DAC was rechecked to evaluate its stability during the XPCS experiments, which always remained constant. In addition, before loaded into the DAC, the as-prepared ribbon sample was initialized by pre-heating into the supercooled liquid region (to eliminate additional stresses induced by fast quenching during sample synthesis). To avoid the influence of shear bands at the sample edges due to mechanical cutting, the x-ray beam was accurately located at the sample center by x-ray absorption scanning at each pressure, which also ensures probing the same sample volume (the sample center) for all the pressure points. The background XPCS signal from the high-pressure environment was carefully checked by shining an x-ray beam on the pressure medium away from the MG sample through the two diamond anvils. No apparent correlation signal was observed at those specific 2θ angles in the background signal (the constant background signal may differ from one specific DAC to another). The data at 0 GPa was collected on the same sample loaded in the DAC before adding the 4:1 M-E pressure medium. More experimental details can be found in the SI Appendix.

***In situ* high-pressure XRD experiments.** *In situ* high-pressure XRD experiments with an x-ray wavelength of 0.3344 Å (37.08 keV) and a beam size of ~3.5 μm × 4.0 μm focused by a Kirkpatrick–Baez mirror were performed at beamline 13-ID-D of Advanced Photon Source (APS), Argonne National Laboratory (ANL), USA. The same sample loading protocol as described above for the XPCS experiment was used for the *in situ* high-pressure XRD experiment. During the experiments, the pressure difference measured before and after each exposure was found to be less than 0.2 GPa. The background patterns were collected at each pressure by shining the x-ray beam off the sample but through the pressure medium and diamond anvils inside the sample chamber. The data at 0 GPa was collected on a sample at ambient conditions outside of the DAC. Raw two-dimensional x-ray diffraction patterns were integrated to obtain the intensities *I(q)* using Dioptas software.

The total scattering factor, $S(q)$ ($q = 4\pi\sin\theta/\lambda$, $\lambda$ is the x-ray wavelength, and 2θ is the angle between the incident and diffraction x-ray beam), and the reduced pair distribution functions, $G(r)$, were derived from $I(q)$ using the PDFgetX3 software package. $S(q)$ is obtained from the coherently scattered intensity $I^{coh}(q)$,

$$S(q) = 1 + \frac{I^{coh}(q) - \langle f^2 \rangle}{\langle f \rangle^2} \tag{2}$$

where $\langle f \rangle = \sum_{i=1}^{n} c_i f_i(q)$ and $\langle f^2 \rangle = \sum_{i=1}^{n} c_i f_{i=1}^2(q)$, in which $c_i$ corresponds to the atomic fraction of the component $i$ having an x-ray atomic scattering factor $f_i(q)$. The *G(r)* function was derived from the Fourier transform of the *S(q)* by the following relation,

$$G(r) = 4\pi r[(\rho(r) - \rho_0] = \frac{2}{\pi} \int_0^\infty q[S(q) - 1] \cdot sin(qr) \, dq \qquad (3)$$

where *ρ(r)* is the pair density function, and *ρ₀* is the average atom number density.

**Acknowledgments**

The authors thank Prof. Jichao Qiao from Northwestern Polytechnical University, Dr. Wenge Yang from HPSTAR, and Prof. Howard Sheng from George Mason University for their helpful discussions, Dr. X.H Chen, Dr. S.Y. Chen, F.J. Lan, and M.B. Sun from HPSTAR for their help with the experimental preparation, Dr. Gaston Garbarino from ESRF for his help with the *in situ* high-pressure XPCS experiments, Dr. Vitali Prakapenka and Dr. Eran Greenberg from University of Chicago for their help with *in situ* high-pressure XRD experiments, and Freyja O'Toole for her language editing. The authors acknowledge financial support from Shanghai Science and Technology Committee, China (No. 22JC1410300) and Shanghai Key Laboratory of Novel Extreme Condition Materials, China (No. 22dz2260800). The *in situ* high-pressure high-energy XPCS experiments were performed at beamline ID10, ESRF. The *in situ* high-pressure XRD experiments were performed at beamlines 13 ID-D of GSECARS, APS, ANL. GSECARS is supported by NSF (EAR-1634415) and DOE (DE-FG02-94ER14466). APS is supported by DOE, Office of Basic Energy Sciences (BES), under Contract DE-AC02-06CH11357. B.R. acknowledges the financial support from the European Research Council (ERC) under the European Union's Horizon 2020 research and innovation program (Grant Agreement No. 948780).

**Figures :**

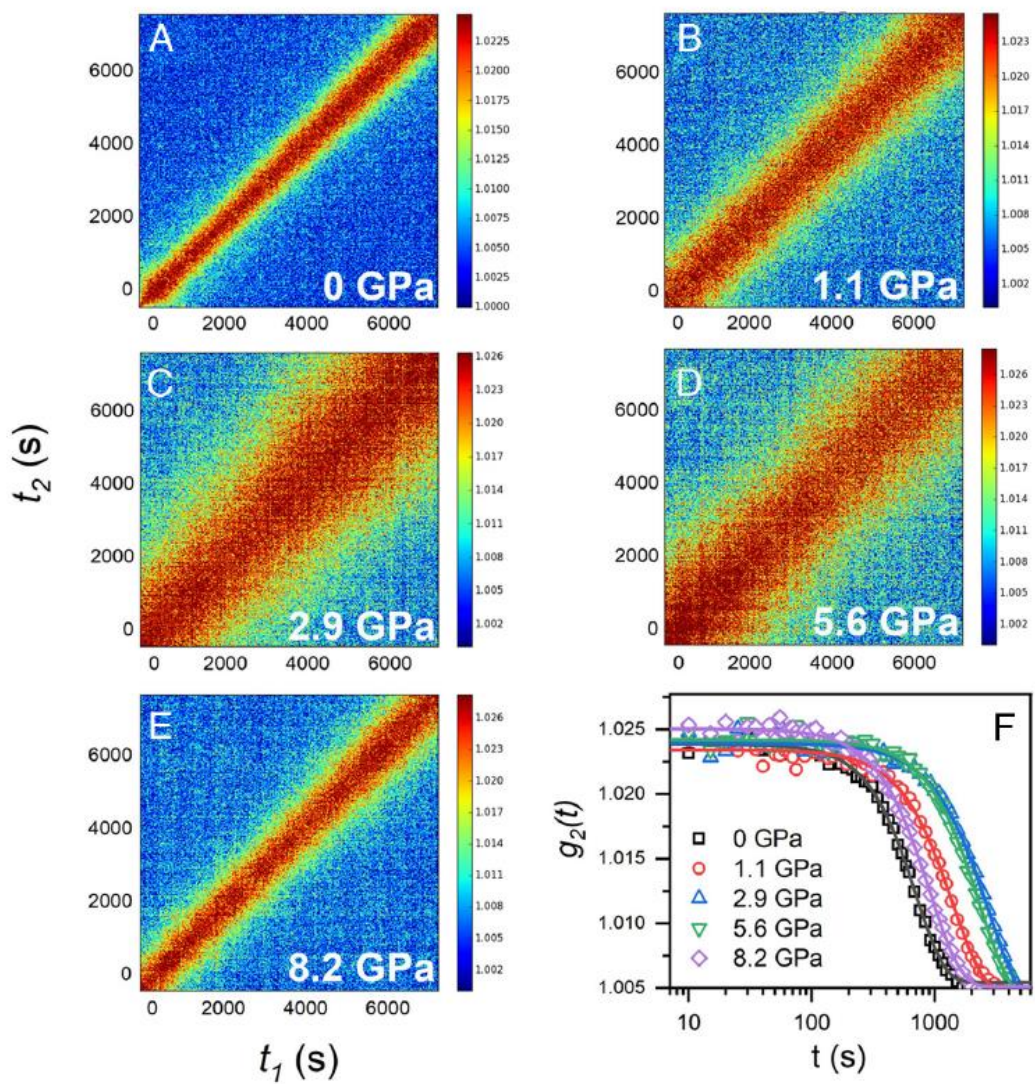

**Fig 1.** Two-time correlation functions of the $Ce_{68}Al_{10}Cu_{20}Co_2$ MG measured at $q_p$ (principal diffraction peak position, as a function of pressure) by XPCS at different pressures during compression. From (a) to (e), the pressures are 0, 1.1, 2.9, 5.6, and 8.2 GPa, respectively. Pressure dependence of intensity correlation functions is shown in (f). The lines are the best fits to a modified KWW model function.

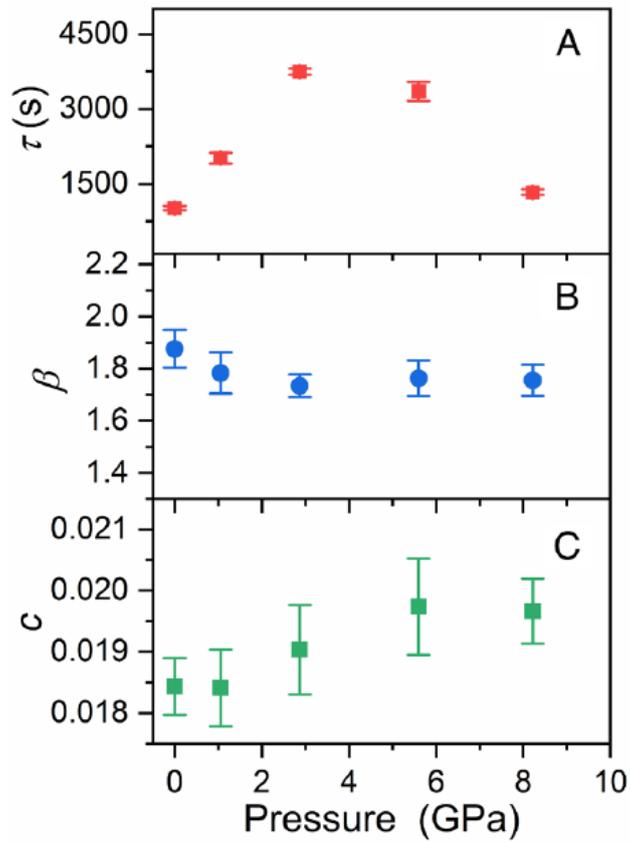

**Fig 2.** The fitting parameters of the intensity correlation functions. Pressure dependence of the relaxation time τ (a), the shape parameter β (b), and the contrast c (c). A clear crossover occurs at ~3 GPa for τ. The decay time is fixed at 1.006 during the fitting of all the data for a proper comparison (see SI Appendix). Vertical error bars are from the KWW model function fitting. Pressure uncertainty is smaller than the symbol size.

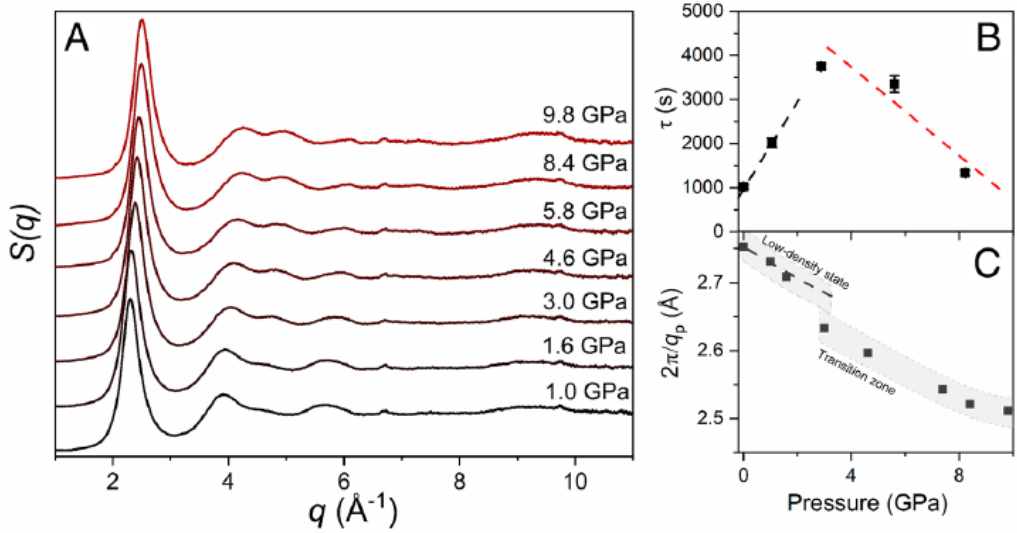

**Fig 3.** Correlation between structural changes revealed by structure factor, $S(q)$, in reciprocal space with relaxation dynamics. (a) $S(q)$ data from 1.0 to 9.8 GPa during compression. A close correlation between the relaxation dynamics and structural evolution is revealed by comparing the pressure dependence of relaxation time $\tau$ measured by XPCS (b) and the inverse of the principal peak position $2\pi/q_p$ in $S(q)$ (c). The data point at 0 GPa is from an ambient condition data without a DAC. The dotted and dashed lines and gray zones are guides to the eye. Error bars in (c) are smaller than the symbol size. It should be noted that due to the discrete data points and the complex pressure dependence in (b) and (c), the exact values of the critical pressures for the changes in (b) and (c) are challenging to be determined.

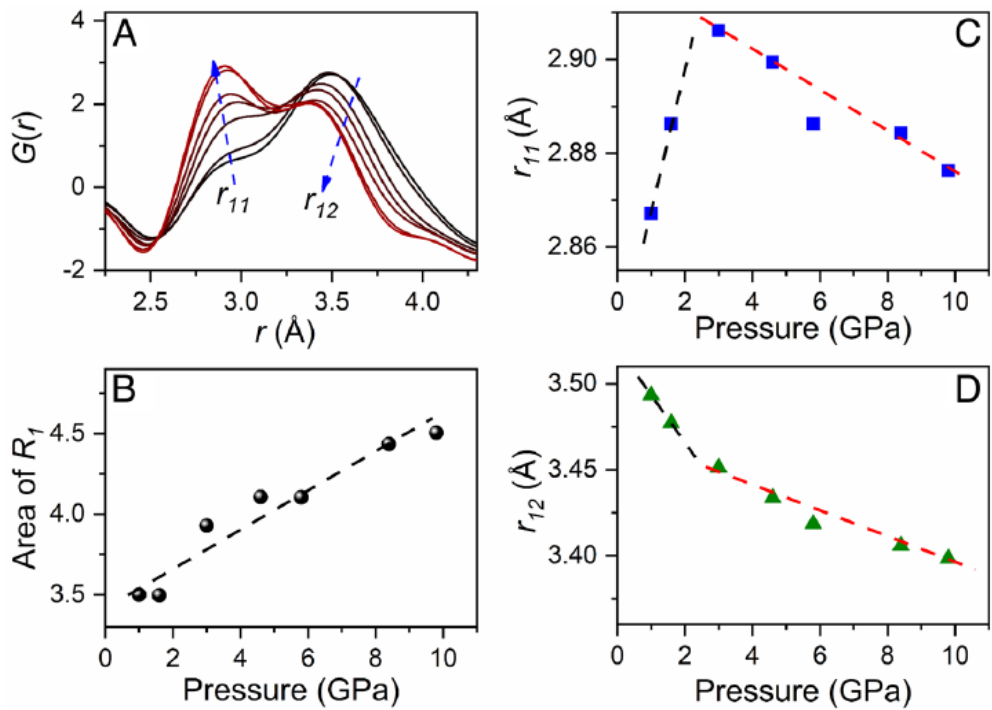

**Fig 4.** Structural changes revealed by reduced PDF, $G(r)$, in real space. (a) The first peak of $G(r)$, $R1$, as a function of pressure from 1.0 to 9.8 GPa during compression. The arrows point to the directions of sub-peak intensity changes. (b) Pressure dependence of the peak area ($A$) of $R1$. Pressure dependence of the first sub-peak position ($r_{11}$) (c) and the second sub-peak position ($r_{12}$) (d) both show a crossover at ~3 GPa. Dashed lines are guides to the eye.